\documentclass[aps,preprint,superscriptaddress]{revtex4}
\usepackage{graphicx}
\usepackage{float}
\usepackage[small,hang,bf]{caption}
\usepackage{subfigure}
\usepackage{bm}    
\usepackage{amsmath}  
\usepackage{amssymb}
\usepackage{times}
\usepackage{color}
\newcommand{\qed}{\nobreak \ifvmode \relax \else
      \ifdim\lastskip<1.5em \hskip-\lastskip
      \hskip1.5em plus0em minus0.5em \fi \nobreak
      \vrule height0.75em width0.5em depth0.25em\fi}

\begin{document}
\title{HILBERT-SPACE PARTITIONING OF THE MOLECULAR ONE-ELECTRON DENSITY MATRIX WITH ORTHOGONAL PROJECTORS}
\author{Diederik Vanfleteren} 
\affiliation{Members of the Ghent Brussels quantum chemistry and molecular modelling alliance.}
\affiliation{Ghent University, Center for Molecular Modeling, 
Technologiepark 903, B-9052 Zwijnaarde, Belgium}
\author{Dimitri Van Neck}
\affiliation{Members of the Ghent Brussels quantum chemistry and molecular modelling alliance.}
\affiliation{Ghent University, Center for Molecular Modeling, 
Technologiepark 903, B-9052 Zwijnaarde, Belgium}
\author{Patrick Bultinck}
\affiliation{Members of the Ghent Brussels quantum chemistry and molecular modelling alliance.}
\affiliation{Ghent University, Department of Inorganic and Physical Chemistry, Krijgslaan 281 (S3), B-9000 Gent, Belgium}
\author{Paul W. Ayers}
\affiliation{McMaster University, Department of Chemistry, Hamilton, Ontario L8S 4M1, Canada}
\author{Michel Waroquier}
\affiliation{Members of the Ghent Brussels quantum chemistry and molecular modelling alliance.}
\affiliation{Ghent University, Center for Molecular Modeling, 
Technologiepark 903, B-9052 Zwijnaarde, Belgium}

\date{\today}
\begin{abstract}
A double-atom partitioning of the molecular one-electron density matrix is used to describe atoms and bonds. All calculations are performed in Hilbert space. The concept of atomic weight functions (familiar from Hirshfeld analysis of the electron density) is extended to atomic weight matrices. These are constructed to be orthogonal projection operators on atomic subspaces, which has significant advantages in the interpretation of the bond contributions. In close analogy to the iterative Hirshfeld procedure, self-consistency is built in at the level of atomic charges and occupancies. The method is applied to a test set of about 67 molecules, representing various types of chemical binding. A close correlation is observed between the atomic charges and the Hirshfeld-I atomic charges. 
\end{abstract}
\maketitle
\section{Introduction}

The concept of Atoms In the Molecule (AIM) has always played a central role in classifying and predicting chemical properties. Because this concept does not naturally show up in molecular orbital (MO) theory, there is a sustained interest in extracting chemical atoms and functional groups from MO-based calculations. Among the most popular techniques are Mulliken population analysis \cite{mulliken1955}, Bader's Quantum Chemical Topology (QCT) \cite{bader1991,bader1994,popelier2000}, the Hirshfeld method in its original \cite{hirshfeld1977} and iterative version \cite{bultinck20071, bultinck20072, bultinck20073, bultinck2009}, the iterative stockholder approach \cite{lillestolen2008}, and Mayer's fuzzy atoms \cite{mayer2004}.

Most methods are restricted to the partitioning of the electron density, but not all properties of a quantum mechanical object can be explicitly expressed in terms of the electron density. A more fundamental approach to the AIM should be based on density matrices \cite{li1986}. Because of the inherent non-locality of the density matrix, it was recently argued that a two-index partitioning into atomic (diagonal) and bond (off-diagonal) contributions, is necessary to guarantee the local nature and transferability of the atoms \cite{mayer2009,vanfleteren2010}. The authors recently introduced such an approach \cite{vanfleteren2010}, but the bond matrices had significant contributions from core electrons and free electron pairs, which somewhat blurred the interpretation in terms of chemical bonding. In this paper we attempt to improve the description of atoms and bonds in a molecule. The interpretive problems are overcome by defining atomic weight matrices to be orthogonal projection operators onto one-electron subspaces assigned to atoms. The proposed method bears a close resemblance to the iterative Hirshfeld procedure, as we introduce the requirement of consistency between weight matrices and atomic contributions. 

\section{Double atom partitioning scheme}
The spin-summed one-electron density matrix (1DM) for a singlet $N$-electron molecular wavefunction $\Psi (\bm{x}_1 ,\dots ,\bm{x}_N )$ is expressed as
\begin{eqnarray}
&&\rho_{ij}=N \int d\bm{r} d\bm{r'} \varphi_{i}(\bm{r})  \varphi_{j}(\bm{r'}) \sum_{\sigma} \\
&&\left[ \int d\bm{x}_{2}...\int d\bm{x}_{N} \Psi^{\dagger} (\bm{r'}\sigma,\bm{x}_{2},...\bm{x}_{N})\Psi^{} (\bm{r}\sigma,\bm{x}_{2},...\bm{x}_{N})\right], \nonumber 
\end{eqnarray}
where the matrix representation in an orthogonal basis set $\varphi_{i}(\bm{r})$ was used. A decomposition of the identity matrix into atomic weight matrices \cite{mayer2005} $(w_{A})_{ij}$
\begin{eqnarray}
\delta_{ij}=\sum_{A}(w_{A})_{ij}, 
\end{eqnarray}
can be inserted on both sides of the molecular one-electron density matrix to arrive at a double-index atomic partitioning
\begin{eqnarray}
\rho=\sum_{AB}\rho_{AB}=\sum_{AB} w_{A}\rho w_{B}. 
\label{partitioning}
\end{eqnarray}
The characteristics of such a partitioning are spatially localized contributions $\rho_{AA}$ belonging to an atom A and some delocalized contributions $(\rho_{AB})_{B \neq A}$ that describe the bonding between atoms A and B \cite{vanfleteren2010}. 

If the atomic weight matrices are chosen as projectors onto orthogonal subspaces, which implies that they are both idempotent and orthogonal
\begin{eqnarray}
w_{A}w_{B}=w_{A}\delta_{AB}, 
\label{idempotent}
\end{eqnarray}
then the bond matrices $(\rho_{AB})_{B \neq A}$ have a zero trace and only the atomic matrices $\rho_{AA}$ will contribute to the number of electrons N in the molecule
\begin{eqnarray}
N&=&Tr(\rho)=\sum_{A}Tr(\rho_{AA}). 
\end{eqnarray}
Note that in general the orthogonality of the weight matrices does not imply a zero overlap between the atomic electron densities $\rho_{AA}(\bm{r},\bm{r'})$ in coordinate space, as occurs in e.g. QCT. 

\section{Constructing orthogonal atomic projectors}

The decomposition of one-electron space according to Eq. (\ref{idempotent}) can be done in myriad ways, so a way must be found that generates a chemically relevant decomposition. We found that, starting from the 1DM's $\rho_{AA}^{(0)}$ for isolated atoms, the following recursive scheme $(i=0,1,...)$
\begin{eqnarray}
\rho^{(i)}&=&\sum_{A}\rho_{AA}^{(i)} \: ; \phantom{....}  w_{A}^{(i)}=(\rho^{(i)})^{-\frac{1}{2}} \rho_{AA}^{(i)} (\rho^{(i)})^{-\frac{1}{2}} \nonumber \\
\rho_{AA}^{(i+1)}&=&w_{A}^{(i)}\rho w_{A}^{(i)}
\label{inner1}
\end{eqnarray}
converges and generates weight matrices $w_{A}^{(\infty)}$ that obey Eq. (\ref{idempotent}). In a forthcoming publication we will analyze this result and introduce alternative recursive schemes, including proofs. Here we just mention three points: (1) it is understood that the molecular 1DM $\rho$ used in Eq. (\ref{inner1}) is positive semidefinite by construction and therefore the same holds for the $\rho_{AA}^{(i)}$ and $\rho^{(i)}$; (2) when the 1DM of the isolated atom $\rho_{AA}^{(0)}$ contains only basis functions centered on atom $A$, the subspaces spanned by the eigenvectors of $\rho_{AA}^{(0)}$ are linearly independent, and as a result the weight matrices $w_{A}^{(0)}$ of the 0th iteration already obey Eq. (\ref{idempotent}); (3) depending on the level of theory for the atomic calculation it is possible that a nullspace is generated for $\rho^{(i)}$, making the inverse square root in Eq. (\ref{inner1}) singular. We therefore replace any eigenvalues of the $\rho_{AA}^{(0)}$ and $\rho$ by a small ($10^{-5}$) positive value. This is sufficient to ensure that all $\rho^{(i)}$ remain nonsingular.   

\section{Self-consistency requirements}
\label{self-consistent-atoms}
 
As in ordinary Hirshfeld, the choice of the starting point (the isolated atom 1DM's $\rho_{AA}^{(0)}$) is crucial, since different results are obtained depending on the charge and electronic state of the isolated atom. Following the ideas behind the iterative Hirshfeld procedure \cite{bultinck20071}, the result can be made independent of the starting point by building in self-consistency through an outer iterative scheme (where Eq. (\ref{inner1}) would represent the inner iterative scheme). Note that the need for a fitted start point

In our simplest implementation (called 'charge equalization') we start from rotationally averaged 1DM's of the neutral isolated atoms. The recursive scheme in Eq. (\ref{inner1}) generates effective electron numbers $N_{A}=Tr(\rho_{AA}^{\infty})$ for the atoms. These can be used to create a rotationally averaged 1DM of the charged isolated atom according to the linear interpolation between integer electron numbers $(k\leq N_{A}<k+1)$
\begin{eqnarray}
\rho_{AA}^{(0)}[N_{A}^{}]&=& \left(k+1-N_{A}^{}\right) \rho_{AA}^{(0)}[k] \nonumber \\&+&\left(N_{A}^{}-k\right)\rho_{AA}^{(0)}[k+1].
\end{eqnarray}
The charged $\rho_{AA}^{(0)}[N_{A}^{}]$ can be used as the next starting point in Eq. (\ref{inner1}) and the whole process is repeated until convergence for the effective electron numbers.

We noticed that, in contrast to the electron density, the 1DM is much more sensitive to a mismatch in the orientation of one-electron orbitals in the molecule and in the isolated atom used to set up the AIM. In some cases, this resulted in rather large AIM charges. In order to solve this problem we also implemented a more sophisticated scheme (called 'population equalization').
The simple implementation of the previous paragraph is followed up to the point where $\rho_{AA}^{(0)}[N_{A}]$ is obtained. We then use the eigenvalue decompositions
\begin{eqnarray}
\rho_{AA}^{(\infty)}&=& \sum_{k} n_{AA,k}^{(\infty)} \varphi_{AA,k}^{(\infty)}\varphi_{AA,k}^{(\infty)} \\
\label{eq1}
\rho_{AA}^{(0)}[N_{A}]&=& \sum_{l} n_{AA,l}^{(0)} \varphi_{AA,l}^{(0)}\varphi_{AA,l}^{(0)}
\label{eq2}
\end{eqnarray}
to generate the new starting point to Eq. (\ref{inner1}). This is given by Eq. (\ref{eq2}), but with the rotationally averaged occupation numbers $n_{AA,l}^{0}$ replaced with the occupation numbers $n_{AA,k}^{(\infty)}$ of the partitioned atom. The correspondence is made on the basis of maximal orbital overlap $|\langle \varphi_{AA,k}^{(\infty)}|\varphi_{AA,l}^{(0)} \rangle|$. By this procedure, the rotationally averaged 1DM's are replaced by 1DM's containing information on the molecular geometry. The whole process is again repeated until convergence.

\section{Results and discussion}
The procedure described in Sec. \ref{self-consistent-atoms} was tested by partitioning the 1DM of a small set of ca. 67 simple molecules with a singlet ground state, representative of a variety of chemical bonds. The set of molecules comprises the species tested in \cite{vanfleteren2010}, supplemented with some extra molecules with relatively high Hirshfeld-I charges like CF$_{4}$ and H$_{2}$SO$_{4}$.  The molecular and atomic 1DM's were calculated at the restricted Hartree-Fock level of theory using the Aug-cc-pVDZ basis set \cite{kendall1992,woon1993,woon2009}. The molecular geometry was taken from a B3LYP \cite{becke1993,lee1988,vosko1980,stephens1994} /cc-pVDZ \cite{woon1993,woon2009,dunning1989} optimization. For CO additional calculations were performed in larger (Aug-cc-pVTZ and Aug-cc-pVQZ) \cite{kendall1992,woon1993,woon2009} basis sets in order to assess basis set convergence.

\subsection{The CO molecule}

\begin{figure}[]
\centering
\includegraphics[width=12cm] {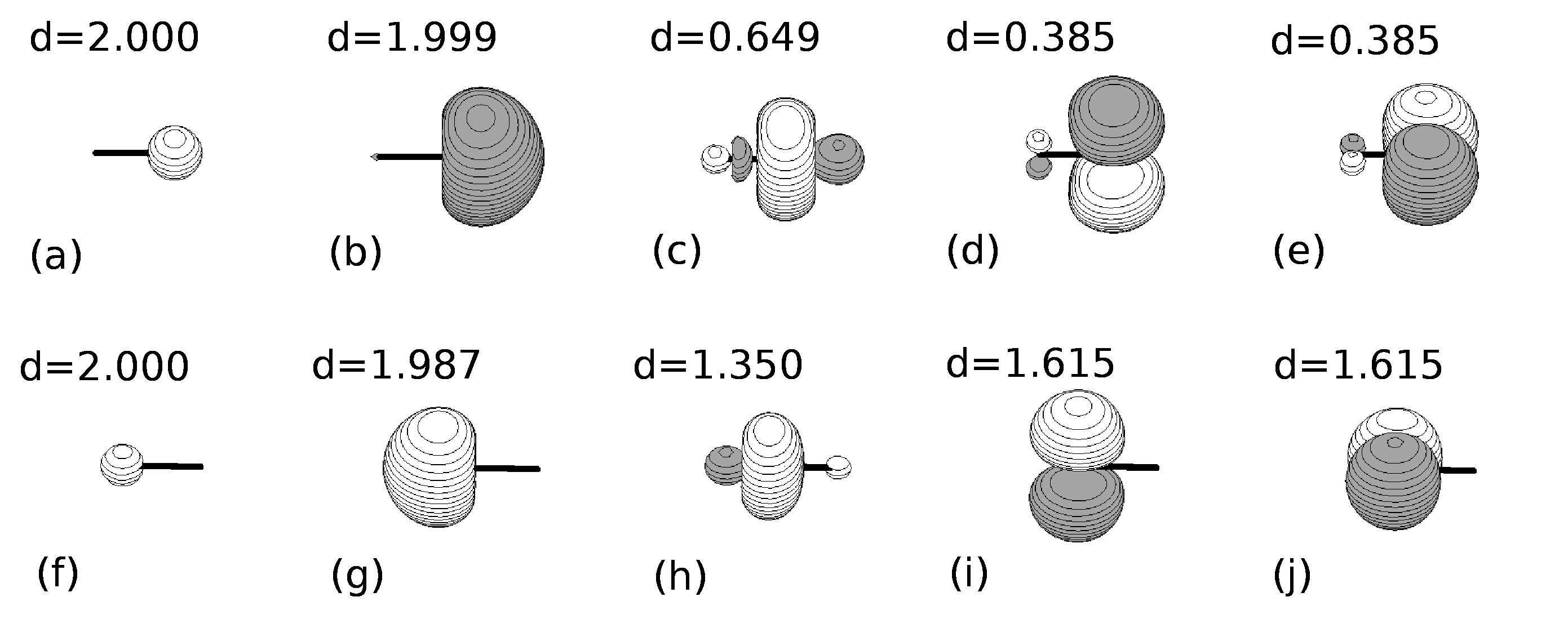}
\caption{The dominant natural orbitals, and the corresponding occupancies of the atomic density matrix of carbon (a-e) and oxygen (f-j) in a CO molecule. For details, see text.
\label{CO_atoms_orbs}}
\end{figure}

Diagonalization of the matrices in Eq. (\ref{partitioning}) leads to natural orbitals and occupancies. Fig. \ref{CO_atoms_orbs} presents the dominant natural orbitals, and the corresponding occupancies, of the atomic density matrix of carbon (a-e) and oxygen (f-j) in a CO molecule using the population equalization scheme. 

Similar to the double-index partitioning  in 3D space based on the use of Hirshfeld-I weights \cite{vanfleteren2010}, the natural orbitals are slightly deformed versions of the typical atomic 1$s$ (a), 2$s$ (b), 2$p_{z}$, 2$p_{x}$ and 2$p_{y}$ (c-d-e) orbitals. However, in contrast to \cite{vanfleteren2010}, the current scheme generates occupancies of the orbitals not involved in bonding (core orbitals or orbitals containing free electron pairs) that are very close to their expected integer value. Also note that the double occupied molecular $\sigma$ -, $\pi_{1}$ and $\pi_{2}$ Hartree-Fock orbitals simply divide their occupancy over the atomic $p_{z}$, $p_{x}$ and $p_{y}$ orbitals. The summed occupancy of (c) and (h) in Fig. \ref{CO_atoms_orbs} is exactly 2. 

\begin{figure}[]
\centering
\includegraphics[width=12cm] {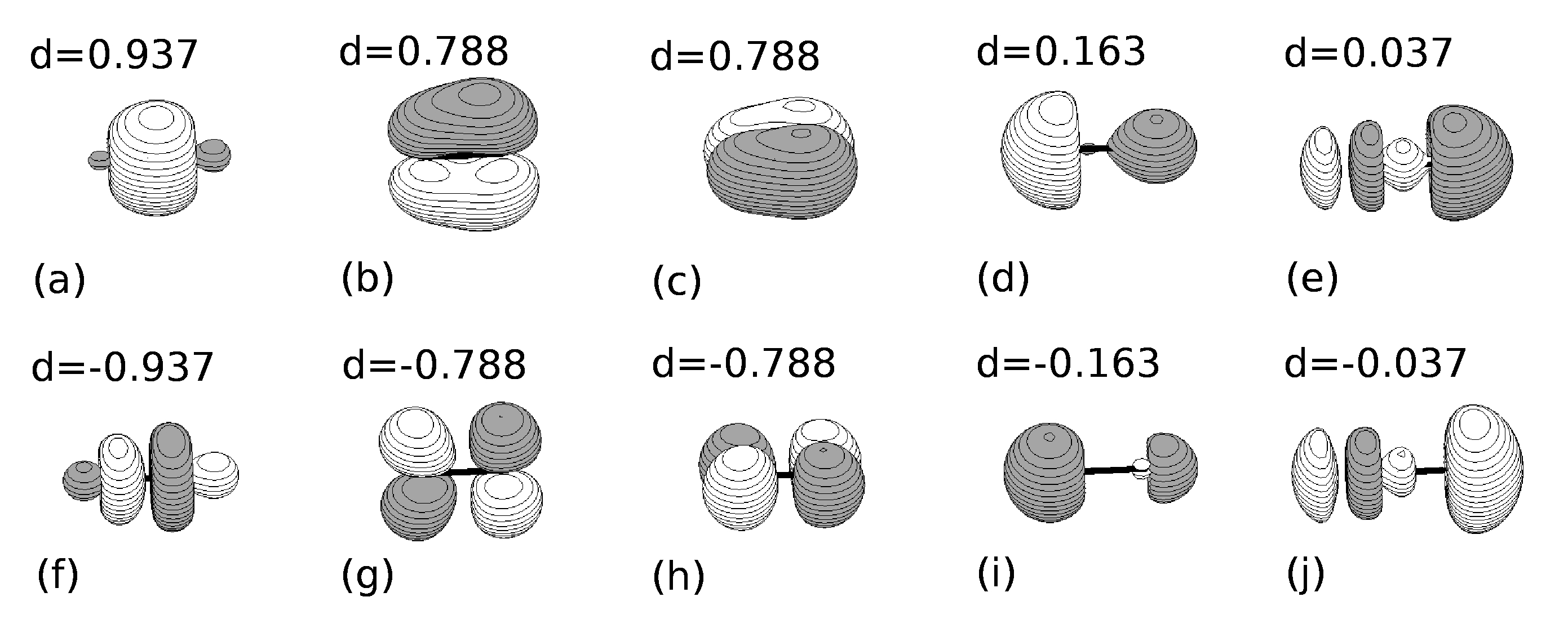}
\caption{The dominant eigenvectors and eigenvalues of (twice) the CO bond matrix in a CO molecule. For details, see text.
\label{CO_bonds_orbs}}
\end{figure}

Fig. \ref{CO_bonds_orbs} depicts the dominant eigenvectors and their eigenvalues of the CO bond matrix $\rho_{AB}$ in the CO molecule. Their shapes match those of the expected bonding and antibonding orbitals, but in contrast to \cite{vanfleteren2010}, the corresponding bonding and antibonding orbitals have exactly opposite eigenvalues by construction. 
The $\sigma*\rightarrow \sigma$, $\pi_{1}* \rightarrow \pi_{1} $ and $\pi_{2}* \rightarrow \pi_{2}$ electron relocations are the main contributions to the bond matrix, though there are some smaller contributions (i $\rightarrow$ d , j $\rightarrow$ e) that have a mainly non-bonding character. Note that we avoid the term "occupancy" for the bond matrices, as the bond matrix reflects a change in density rather then a density itself. 

Apart from the dominant contributions shown in Figs.~\ref{CO_atoms_orbs}-\ref{CO_bonds_orbs}, various orbitals with much smaller eigenvalues are also present (as the Aug-cc-pVDZ molecular basis set has 46 basis functions). A complete overview is given in Table \ref{tab2} for both population equalization and charge equalization schemes (bracketed values). Note that for the atomic density matrices, apart from the five dominant natural orbitals, only two more have small ($\sim 10^{-2}-10^{-3}$) populations, while the remainder have vanishing $(< 10^{-13})$ populations. For the bond matrix, apart from the ten dominant orbitals, only four more have small eigenvalues, as was already noted in \cite{vanfleteren2010}.

\begin{table}[]
\centering
\begin{tabular}{ l l l l l l}
\hline 
     &  C , C        &  O , O          &  C , O  + O , C                   \\  
\hline
(a) & 2.000 ( 2.000 ) & 2.000 ( 2.000 ) & $\pm$ 0.937 ( $\pm$ 0.919 ) \\
(b) & 1.999 ( 1.988 ) & 1.987 ( 1.992 ) & $\pm$ 0.788 ( $\pm$ 0.799 ) \\
(c) & 0.649 ( 0.605 ) & 1.615 ( 1.601 ) & $\pm$ 0.788 ( $\pm$ 0.799 ) \\
(d) & 0.385 ( 0.399 ) & 1.615 ( 1.601 ) & $\pm$ 0.163 ( $\pm$ 0.154 ) \\
(e) & 0.385 ( 0.399 ) & 1.350 ( 1.395 ) & $\pm$ 0.037 ( $\pm$ 0.128 ) \\
(f) & 0.013 ( 0.008 ) & 0.001 ( 0.012 ) & $\pm$ 0.014 ( $\pm$ 0.009 ) \\
(g) & 0.000 ( 0.000 ) & 0.000 ( 0.000 ) & $\pm$ 0.011 ( $\pm$ 0.007 ) \\
(...)        & $<10^{-13}$   & $<10^{-13}$  & $-10^{-13}><10^{-13}$     \\
\hline
Sum & 5.432 ( 5.400 ) & 8.568 ( 8.600 ) & 2.738 ( 2.816 )  \\
\hline 
\end{tabular} 
\caption{All natural populations in the atomic density matrices (CC and OO) and eigenvalues of the bond matrix (CO) in a CO molecule. Non-bracketed and bracketed values belong to the population equalization scheme and the charge equalization scheme respectively. For details, see text.} \label{tab2}
\end{table}

In Table~\ref{tabCOnw} the stability of the proposed density matrix partitioning is examined. The results seem to be quite stable with respect to basis set size, with differences for the traces of atomic density matrices less than 0.01 going from DZ to TZ, and less than 0.001 going from TZ to QZ. For the positive and negative components of the bond matrices there are fluctuations of about 0.02 from DZ to TZ and from TZ to QZ. We checked that the stability holds even at the level of the individual orbital eigenvalues in Table \ref{tab2}. In the charge equalization scheme, all eigenvalues differ less than 0.003 going from DZ to TZ, and less than 0.002 from TZ to QZ. In the population equalization scheme, the atomic matrices behave similarly (maximal deviation of 0.005 from DZ to TZ and of 0.003 from TZ to QZ), but the bond matrix eigenvalues deviate slightly more (maximal deviation of 0.015 from DZ to TZ and of 0.022 from TZ to QZ). We also checked that for almost all the orbitals in Figs. Figs.~\ref{CO_atoms_orbs}-\ref{CO_bonds_orbs}, the shapes are visually indistinguishable when changing basis set size from DZ to QZ. The exceptions are the bond matrix orbitals (e) and (j), corresponding to small (in absolute value) eigenvalues, where a somewhat more distorted shape is observed.

\begin{table}[]
\centering
\begin{tabular}{|l| l| l| l| l| l|}
\hline 
CO & Aug-cc-pVDZ & Aug-cc-pVTZ & Aug-cc-pVQZ  \\
\hline 
A , B & Tr($\rho_{A,B}$) & Tr($\rho_{A,B}$) & Tr($\rho_{A,B}$) \\
\hline
C , C              &  5.432 ( 5.400) &  5.425 ( 5.393) &  5.426 (5.394)\\
O , O              &  8.568 ( 8.600) &  8.575 ( 8.607) &  8.574 (8.606)\\
C , O + O , C      &  $\pm$ 2.738 ( $\pm$2.816) &  $\pm$ 2.756 ( $\pm$ 2.811) &  $\pm$ 2.731 ( $\pm$ 2.814)\\
\hline 
\end{tabular} 
\caption{Basis set convergence of the summed positive and negative eigenvalues in the atomic and bond matrices for CO. Non-bracketed and bracketed values belong to the population equalization scheme and the charge equalization scheme respectively. For details, see text. \label{tabCOnw}}
\end{table}

\subsection{Evaluation of the atomic charges}

\begin{figure}[]
\centering
\includegraphics[width=7cm] {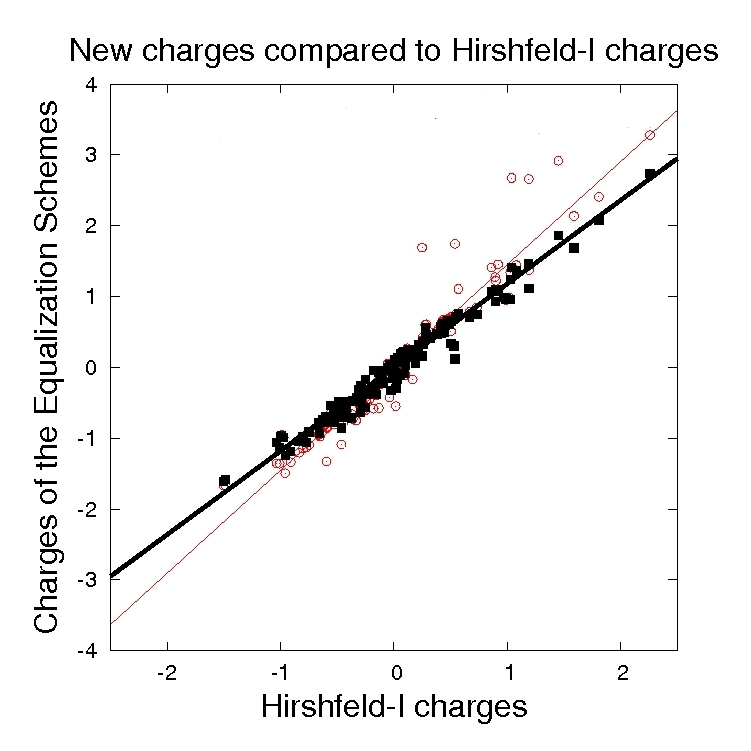}
\caption{Comparison of the Hirshfeld-I charges with those generated by the charge equalization scheme (red circles) and the population equalization scheme (black squares). For details, see text.
\label{eval-charges}}
\end{figure}

Fig. \ref{eval-charges} displays the correlation between the charges obtained with the Hirshfeld-I method \cite{bultinck20071} (known to be a reliable AIM method) and the charges resulting from either the charge equalization scheme or the population equalization scheme applied to the entire set of 67 molecules. Both schemes show a strong linear correlation with Hirshfeld-I. It is remarkable that the linear correlation is more satisfying for the population equalization scheme (R$^{2}$=0.96, slope=1.18) than when only the atomic charges are made self-consistent (R$^{2}$=0.90, slope = 1.45).  As mentioned, this is related to the inadequacy of the rotationally averaged 1DM's used as starting point in the charge equalization scheme. In some cases, the latter leads to unexpectedly high charges. This problem seems to be solved using the population equalization procedure, which is more adapted to the molecular geometry.

\section{Conclusions}
We have implemented a two-index partitioning of the molecular density matrix into atomic and bond contributions using atomic weight matrices that are orthogonal projection operators in one-electron Hilbert space. The method is highly efficient in terms of computation time since no 3D integrals need to be computed. The resulting decomposition provides a rather physical description of the adaptations made by atoms in a molecule and the deformation caused by the bonding process. The bond matrices are traceless, i.e. the electrons are all in the atomic contributions and not in the bond matrices. Core orbitals and free electron pairs are fully assigned to the atoms. The trial atoms and the AIM atoms are required to have equal charges or equal orbital populations. The latter approach leads to a better correlation of the atomic charges with the Hirshfeld-I atomic charges.

%
%
   \bibliographystyle{unsrt}

\end{document}